\documentclass[prb,twocolumn,amsmath,amssymb,aps]{revtex4-2}

\usepackage{bm}% bold math
\usepackage{hyperref}% add hypertext capabilities
\usepackage[svgnames]{xcolor} % for different colorfonts (comments)
\usepackage{graphicx}
\usepackage[english]{babel}
\usepackage{comment}
\usepackage{xspace}  % To add/remove a space after commands
\usepackage[normalem]{ulem}  % \sout

\makeatletter
\let\ORIbbl@fixname\bbl@fixname
\def\bbl@fixname#1{%
  \@ifundefined{languagealias@\expandafter\string#1}
    {\ORIbbl@fixname#1}
    {\edef\languagename{\@nameuse{languagealias@#1}}}%
}
\newcommand{\definelanguagealias}[2]{%
  \@namedef{languagealias@#1}{#2}%
}
\makeatother
\definelanguagealias{en}{english}

\usepackage{scalerel,stackengine}
\stackMath
\newcommand\reallywidehat[1]{%
\savestack{\tmpbox}{\stretchto{%
  \scaleto{%
    \scalerel*[\widthof{\ensuremath{#1}}]{\kern-.6pt\bigwedge\kern-.6pt}%
    {\rule[-\textheight/2]{1ex}{\textheight}}%WIDTH-LIMITED BIG WEDGE
  }{\textheight}% 
}{0.5ex}}%
\stackon[1pt]{#1}{\tmpbox}%
}

\newcommand*{\angstrom}{\text{\normalfont\AA}}
\newcommand*{\cgp}{c_{\mathrm{0}}}
\newcommand*{\xigp}{\xi_{\mathrm{0}}}
\newcommand*{\fchi}{f_{\chi,\gamma}}
\newcommand*{\Vi}{V_{\mathrm{I}}}
\newcommand*{\Vdisk}{V_{\mathrm{obs}}}
\newcommand*{\Vxy}{\Vi(\bm{x} - \bm{y})}
\newcommand*{\krot}{k_{\mathrm{rot}}}
\newcommand*{\He}{$^4$He\xspace}
\newcommand*{\ML}{M_L\xspace}
\newcommand*{\vonKarman}{B\'enard-von K\'arm\'an\xspace}

\begin{document}

\title{Critical velocity for vortex nucleation and roton emission in a generalized model for superfluids}

\author{Nicol\'as P. M\"uller}
\author{Giorgio Krstulovic}
\affiliation{%
  % http://univ-cotedazur.fr/fr/recherche/la-signature-scientifique-duca
  Université Côte d'Azur, Observatoire de la Côte d'Azur, CNRS,
  Laboratoire Lagrange,  Boulevard de l'Observatoire CS 34229 - F 06304 NICE Cedex 4, France
}

% \pacs{}
% \keywords{A; B; C...}

\begin{abstract}
  We study numerically the process of vortex nucleation at the wake of a moving object in
  superfluids using a generalized and non-local Gross-Pitaevskii model. The non-local potential is
  set to reproduce the roton minimum present in the excitation spectrum of superfluid helium. By
  applying numerically a Newton–Raphson method we determine the bifurcation diagram for different
  types of non-linearities and object sizes which allow for determining the corresponding critical
  velocities. In the case of a non-local potential, we observe that for small object sizes the
  critical velocity is simply determined by the Landau criterion for superfluidity whereas for large
  objects there is little difference between all models studied. Finally, we study dynamically in
  two and three dimensions how rotons and vortices are excited in the non-local model of superfluid.
\end{abstract}

\maketitle

\section{Introduction}

One of the most interesting features of superfluids is their total absence of viscosity. 
This means that a particle traveling in a superfluid experiences no drag force and moves freely with no friction.
However, it took no long to Landau to realize that if a moving impurity exceeds a certain velocity, known as Landau's critical velocity \cite{Pitaevskii2016} 
\begin{equation}
        v_L = \min_{\bm{k}} \frac{\omega(\bm{k})}{|\bm{k}|},
        \label{eq:landau}
\end{equation}
with $\omega(\bm{k})$ the dispersion relation of the superfluid and $\bm{k}$ the wave vector, 
it generates the spontaneous creation of elementary excitations that act as a dissipative mechanism on the impurity. This is known as Landau's criterion for superfluidity. %, given that below this critical velocity there is no dissipation.
In a non-interacting Bose-Einstein condensate (BEC), the dispersion relation is proportional to $k^2$ so Landau's velocity is zero and superfluidity can not take place. In a weakly interacting BEC, the system follows the Bogoliubov dispersion relation \cite{Dalfovo1999} and Landau's velocity is given by the speed of sound of the superfluid $v_{\mathrm{L}}=c$, while in superfluid \He, Landau's critical velocity is smaller than the speed of sound as a consequence of the well-known roton minimum appearing in its excitation spectrum \cite{DonnellyRJ1991, Godfrin2021}.

In classical compressible fluids, velocities above the speed of sound would lead to the formation of shock waves \cite{Landau1987}. However, shock waves in superfluids are suppressed due to the dispersive nature of the system. Instead, these structures are replaced by the nucleation of vortices.
In the early 90s it was first observed numerically in weakly interacting BECs that a particle traveling through a superfluid may experience a drag force if it exceeds a critical velocity $v_c$ \cite{Frisch1992}, nucleating vortices at its wake. This critical velocity was found to be smaller than the speed of sound. The reason of this is that the local velocity of the flow exceeds the speed of sound somewhere around the surface of the obstacle.
Since then, several efforts were carried out to provide a better description on the mechanisms of vortex nucleation, in particular, in the determination of the critical velocity of superfluids and its dependence with the size of the moving obstacle \cite{HuepeCristian1997,Winiecki1999, Winiecki2000, Huepe2000, Pham2005, Rica2001, Villois2018}. 
The nucleation of vortices is a process that takes place in different quantum flows, like BECs \cite{Raman1999, Kwon2016}, superfluid of light \cite{Eloy2021, Carusotto2013}, and superfluid \He \cite{Efimov2010}.
Numerical simulations in models of BECs and dipolar BECs showed that the obstacle can create regular or irregular vortex patterns at its  wake, in particular the creation of a \vonKarman vortex street \cite{Sasaki2010, Stagg2014, Reeves2015, Xi2021}. 
%In later numerical works it was shown that this velocity depends on the size of the particle \cite{HuepeCristian1997, Winiecki1999, Winiecki2000, Huepe2000, Pham2005, Sasaki2010, Villois2018}. 

Understanding the process of vortex nucleation is very important for its practical applications.
For instance, it can be used as a mean of injecting vortices and energy into a system as in grid turbulence \cite{Salort2010, Krstulovic2016} and is also a relevant process on the study of lift force of a flow around an airfoil \cite{Musser2019}.
The study of vortices in superfluid \He presents some difficulties given that there is not a simple microscopic description of it. However, it is possible to study some of its phenomenology assuming a non-local interaction between the bosons constituting the superfluid \cite{Berloff1999,Reneuve2018,Muller2020,Berloff2000}. In this framework, a moving obstacle is allowed to emit some density excitations known as rotons \cite{Berloff2000,Pomeau1993,Kolomeisky2021}.

In this work, we focus on the determination of the critical velocity for the nucleation of vortices in different zero-temperature models for superfluids. In particular, we study a model that better describes weakly interacting BECs where compressibility effects can vary, and a model that incorporates a roton minimum in the excitation spectrum.
In particular, we show the differences between the vortex nucleation and roton creation processes.
In section \ref{sec:model} we introduce the different models used in this work and, in particular, with the presence of a moving obstacle. In section \ref{sec:nucleation} we present the results obtained on the study of the vortex nucleation in these different models both in the stationary and dynamical regimes in two and three dimensions and, finally, in section \ref{sec:conclusions} we present our conclusions.

\section{Model for superfluid \He}
\label{sec:model}

A superfluid at zero temperature constituted by bosons of mass $m$ can be described by the generalized Gross-Pitaevskii (gGP) equation \cite{Berloff1999, Berloff2014, Muller2020}
\begin{equation}
\begin{split}
        i \hbar \frac{\partial\psi}{\partial t} = &-\frac{\hbar^2}{2m}\nabla^2 \psi - \mu(1+\chi) \psi \\
                                                  &+ g  \left(\int \Vi(\bm{x}-\bm{y})|\psi(\bm{y})|^2 \mathrm{d}^3y \right) \psi + g\chi\frac{|\psi|^{2(1+\gamma)}}{n_0^{\gamma}}\psi,  
\end{split}
\label{eq:gGP_hbar}
\end{equation}
where $\psi$ is the macroscopic wave function of the condensate, $\mu$ the chemical potential, $g=4\pi \hbar^2 a_s/m$ the coupling constant fixed by the $s$-wave scattering length $a_s$, and $n_0$ the particles density of the ground state. The last term is a high-order correction of the mean field approximation, with $\chi$ and $\gamma$ two dimensionless parameters corresponding to its amplitude and order, respectively. The chemical potential has been renormalized so that $|\psi_0|^2 = n_0$ remains the ground state of the system. The interaction potential between bosons $\Vi$ is normalized such that $\int \Vi(\bm{x}) \mathrm{d}^3x = 1$. Note that by choosing a $\delta$-function interaction potential $\Vi(\bm{x}-\bm{y}) = \delta(\bm{x} - \bm{y})$, and setting $\chi=0$, one recovers the standard Gross-Pitaevskii (GP) equation \cite{Pitaevskii2016}. We will refer as the local gGP model the case where the interaction potential is a $\delta$-function, but the beyond mean field corrections are not neglected, i. e. $\chi \neq 0$, the local gGP model.

Perturbing the system around the ground state recovers the generalized Bogoliubov dispersion relation of the system
\begin{equation}
        \omega_B(\bm{k}) = c k\sqrt{\frac{\xi^2k^2}{2} + \frac{\hat{\Vi}(\bm{k}) + \chi(\gamma + 1)}{1 + \chi(\gamma+1)}},
\label{eq:bogoliubov}
\end{equation}
where $k$ is the wave number of the perturbation and $\hat{\Vi} = \int e^{i\bm{k}\cdot\bm{r}} \Vi(\bm{r}) \mathrm{d}^3 r$ is the Fourier transform of the interaction potential normalized such that $\hat{\Vi}(k=0)=1$. The speed of sound and healing length of the system are respectively given by
\begin{align}
c &= \cgp \sqrt{\fchi}, \label{eq:c}\\
\xi &= \frac{\xigp}{\sqrt{\fchi}}, \label{eq:healing}
\end{align}
with $\cgp=\sqrt{gn_0/m}$ and $\xigp=\hbar/\sqrt{2mgn_0}$ the speed of sound and healing length of the standard GP model, respectively. The factor $\fchi = 1+\chi(\gamma+1)$ is a rescaling parameter of the system. 
Larger values of $\chi$ or $\gamma$ correspond to stronger interactions between bosons, thus making the fluid more incompressible. As a consequence, the speed of sound increases at the same rate as the healing length decreases.
Note that the product between $c$ and $\xi$ is independent of the high-order corrections and is associated with the quanta of circulation $\kappa = c\xi 2\pi\sqrt{2} = h/m$ that depends only on the mass of the bosons constituting the superfluid. 

The gGP model \eqref{eq:gGP_hbar} can be rewritten in terms of the relevant parameters of the system as 
\begin{equation}
\begin{split}
        \frac{\partial\psi}{\partial t} = &-i\frac{c}{\xi\sqrt{2}\fchi} \bigg[ -\fchi\xi^2\nabla^2\psi - (1+\chi) \psi \\
                                          &+\frac{1}{n_0}  (\Vi*|\psi|^2) \psi + \frac{\chi}{n_0^{\gamma+1}}|\psi|^{2(\gamma+1)}\psi  \bigg].
\end{split}
\label{eq:gGP}
\end{equation}
This generalized model can be used to provide a better phenomenological description of different systems like superfluid \He \cite{Muller2020}, dipolar gases \cite{Lahaye2009} or even the supersolid state of matter \cite{Gallemi2020a}. In the particular case of superfluid \He, the following isotropic potential \cite{Berloff2014, Reneuve2018}  
\begin{equation}
\hat{\Vi}(\bm{k}) = \left[1 - V_1 \left(\frac{k}{k_{\mathrm{rot}}}\right)^2 + V_2 \left(\frac{k}{k_{\mathrm{rot}}}\right)^4\right]\exp\left(-\frac{k^2}{2k_{\mathrm{rot}}^2}\right),
\label{eq:potential}
\end{equation}
can reproduce the excitation spectrum observed experimentally \cite{Donnelly1998, Godfrin2021}. Here $\krot=2\pi/a_{\mathrm{rot}}$ is the wave number associated to the roton minimum length scale of \He $a_{\mathrm{rot}} = 3.26$ $\angstrom$ and together with the dimensionless parameters $V_1$ and $V_2$ are determined to mimic its experimental dispersion relation \cite{Donnelly1998}. 
In this work, this fit was done by considering that equation \eqref{eq:gGP} is written in terms of the healing length of \He $\xi=0.8$ $\angstrom$ and the turnover time at small scales $\tau=\xi/c=3.36\times10^{-13}$ s, being the speed of sound in \He $c=238$ ms$^{-1}$. Using this system of units it is possible to determine the values of $V_1$, $V_2$ and $k_{\mathrm{rot}}$ to recover the roton minimum in the excitation spectrum \cite{Muller2020}. The beyond mean field correction was implemented to avoid the development of instabilities of wave numbers close to the roton minimum \cite{Reneuve2018}. 
In the following sections, all simulations with a non-local interaction were done with $\gamma=2.8$, $\chi=0.1$, $V_1=4.54$, $V_2=0.01$ and $\krot\xi=1.638$. This particular choice of $\gamma$ is set so that the long-wavelength sound waves are proportional to $\rho^{2.8}$ according to experiments \cite{Berloff1999, Brooks1977}. 

\subsection{Superfluid with a moving obstacle}
\label{subsec:moving}

In superfluid \He, Landau's critical velocity is determined by the roton minimum in the excitation spectrum and is associated with the emission of density fluctuations. 
In the case of an obstacle moving with a velocity $\bm{U} = U\hat{y}$, assuming energy and momentum conservation, Landau's criterion for superfluidity can be rewritten as \cite{Kolomeisky2021}
\begin{equation}
        \bm{k}\cdot\bm{v} - \omega(\bm{k}) = k_y U - \omega(\bm{k}) = 0,
        \label{eq:landau_vy}
\end{equation}
showing that there is some anisotropy and a range of excited wave numbers. 

We can describe an obstacle moving in a superfluid with a Gaussian potential $\Vdisk(\bm{r} - \bm{U}t) = V_0 e^{-\frac{1}{2}\frac{|\bm{r} - \bm{U}t|^2}{\Delta^2}}$ that describes a disk (sphere) in two (three) dimensions. The size of the obstacle in the Thomas-Fermi approximation is determined by $\Delta = D/(2\sqrt{2\mathrm{log}(V_0)})$ with $D$ its diameter. The amplitude of the potential is chosen as $V_0 \gg 1$ so that the obstacle completely depletes the superfluid. Thus, the equation of motion of a superfluid with a moving obstacle becomes
\begin{equation}
\begin{split}
        \frac{\partial\psi}{\partial t} = &-i\frac{c}{\xi\sqrt{2}\fchi} \bigg\{ -\fchi\xi^2\nabla^2 - [1+\chi - \Vdisk(\bm{r}-\bm{U}t)] \\
                                          &+\frac{1}{n_0}  (\Vi*|\psi|^2) + \frac{\chi}{n_0^{\gamma+1}}|\psi|^{2(\gamma+1)} \bigg\}\psi,
\end{split}
\label{eq:gGP_disk}
\end{equation}
with the total energy of the system 
\begin{equation}
\begin{split}
        \mathcal{E} =& \frac{c^2}{n_0 V\fchi} \int \bigg\{\fchi\xi^2 |\bm{\nabla}\psi|^2 + \frac{|\psi|^2}{2n_0}(\Vi*|\psi|^2) - \\
                     & \left[1 + \chi - \Vdisk(\bm{r}-\bm{U}t)\right]|\psi|^2 + \frac{\chi |\psi|^{2(\gamma+2)}}{n_0^{\gamma+1}(\gamma+2)}\bigg\}\mathrm{d}^3r.
\end{split}
\label{eq:Enucleation}
\end{equation}

To determine the critical velocity of the superfluid, it is convenient to study solutions of the system that are stationary in the frame of reference of the moving particle \cite{HuepeCristian1997,Pham2005}.
To do this, we look for steady solutions of the wave function of the form $\psi(\bm{r},t) = \Psi(\bm{r}-\bm{U}t) = \Psi(\tilde{\bm{r}})$ with boundary conditions such that $\psi \xrightarrow{r \rightarrow \infty} \sqrt{n_0}$. 
The equation obtained after performing this transformation is
\begin{equation}
\begin{split}
        \bm{U}\cdot \bm{\nabla}_{\tilde{\bm{r}}}\Psi = &i\frac{c}{\xi\sqrt{2}\fchi} \bigg\{ -\fchi\xi^2\nabla^2 - [1+\chi - \Vdisk(\tilde{\bm{r}})]  \\
                                          &+\frac{1}{n_0}  (\Vi*|\Psi|^2)  + \frac{\chi}{n_0^{\gamma+1}}|\Psi|^{2(\gamma+1)}  \bigg\}\Psi.
\end{split}
\label{eq:gGP_steady}
\end{equation}

\section{Vortex nucleation}
\label{sec:nucleation}

In this section, we study the different dynamics of an object moving at a constant velocity $\bm{U} = U \hat{y}$ in a superfluid at rest. We determine the critical velocity $U_c$ or critical Mach number $M_c = U_c / c$ of the system for different diameters $D$ of the disk, above which it starts nucleating vortices. 
To do this, we perform two-dimensional numerical simulations with periodic boundary conditions of the 
gGP model with a moving particle \eqref{eq:gGP_disk}. 
In all cases we solve the system with a spatial resolution $\Delta x = \xi$ in a squared domain with a size $L > 5 D$ to minimize spurious effects that may surge as a consequence of periodicity, using a number of collocation points that go from $512^2$ to $2048^2$. 
We study the differences of the phenomenon of vortex nucleation for the standard GP model, the local gGP \eqref{eq:gGP_disk} for different values of $\chi$ and $\gamma$ and the non-local gGP with the interaction potential \eqref{eq:potential} that supports roton excitations.

\subsection{Critical velocity in the local gGP model}
\label{subsec:stationary}

\begin{figure}[tpb]
  \centering
  \includegraphics[width=1\linewidth,trim={0 0cm 0 0cm},clip]{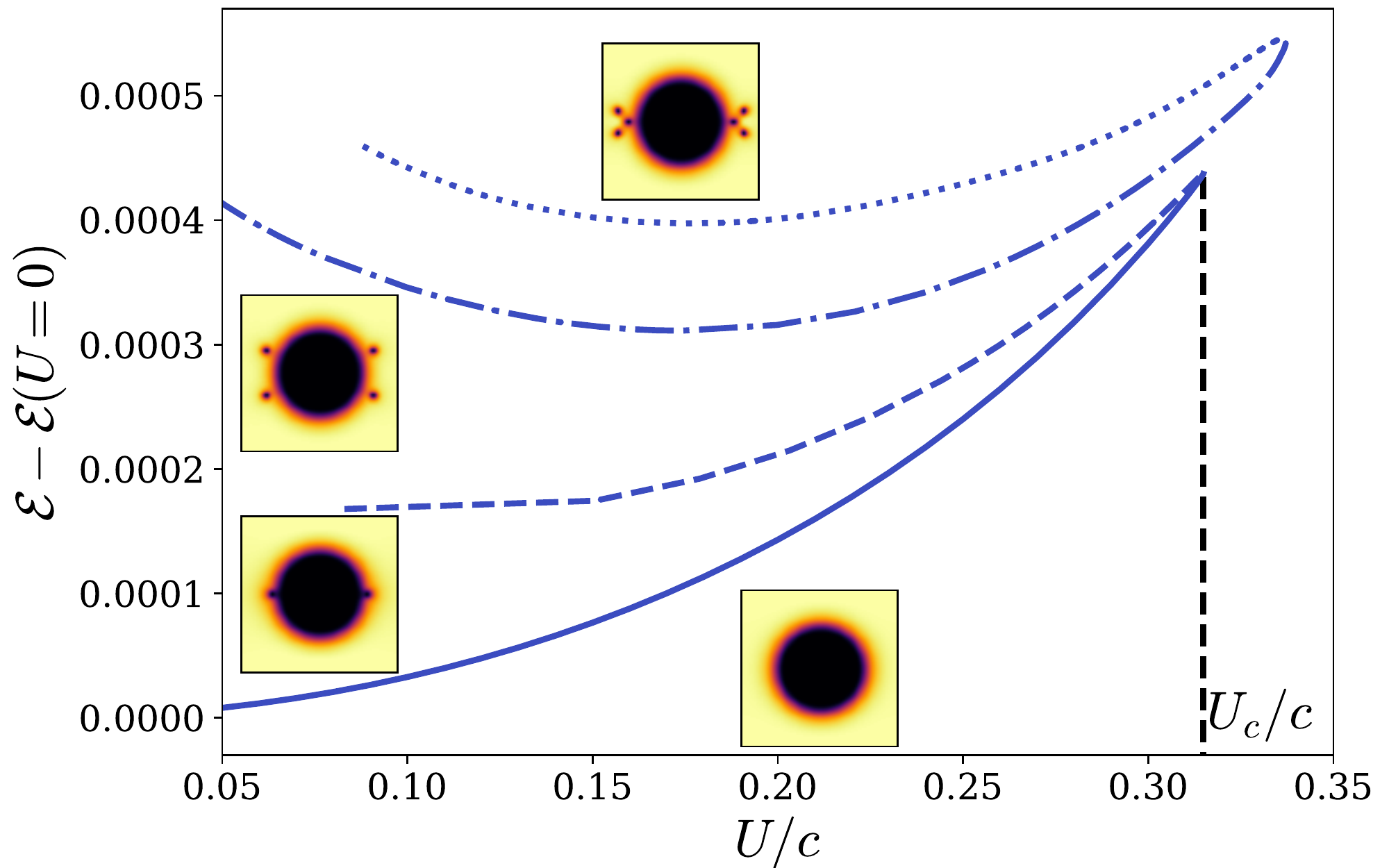}
  \caption{Bifurcation diagram of the energy of stationary solutions in the standard GP model with a disk diameter $D=40 \xi$ moving at different velocities $U$. The stable branch (solid line) and the two vortices (dashed line), four vortices (dot-dashed line) and six vortices (dotted line) unstable branches are shown. The insets show the density fields of the different branches. Dark colors correspond to regions where the density vanishes.}
  \label{fig:branches_gp}
\end{figure}%

As discussed in section \ref{subsec:moving}, the determination of the critical velocity can be done by studying the stationary solutions of the system. 
A superfluid with a moving obstacle counts with different sets of steady solutions, some of them stable and some others unstable \cite{HuepeCristian1997, Pham2005}. The stable stationary solutions of the system can be obtained by solving the imaginary time gGP model, i.e. replacing $t \rightarrow -it$ in Eq.~\eqref{eq:gGP_steady}. 
However, this method only recovers states with minimal energy, that is, it can only be used to recover stable stationary solutions. Therefore, we implement a Newton-Raphson method to be also able to obtain unstable stationary solutions of the system (see Appendix \ref{sec:newton} for details). 

Figure \ref{fig:branches_gp} shows different energy branches $\mathcal{E}-\mathcal{E}(U=0)$ of stationary solutions obtained as the Mach number of the disk $M=U/c$ varies. Each of these values was obtained using a  Newton–Raphson method to solve the standard GP (local interaction potential with $\chi=0$) and with a disk diameter $D = 40 \xi$. 
The other energy branches correspond to unstable solutions in which two (dashed line), four (dot-dashed) or six (dotted) vortices are nucleated. The time evolution of each of these solutions is stationary in the frame of reference of the moving disk, i.e. the number of vortices in the system will not change.
The interesting aspect of the bifurcation diagram of the system is that it provides a way to determine the critical Mach of the superfluid $M_c$ for a particular disk size $D$. 
Such value corresponds to the Mach number where the stable and unstable branches merge together, being in this case $M_c \approx 0.315$. 
Beyond this critical velocity there is no stationary solution, meaning that the disk would nucleate vortices and experience some drag force.
%, reducing its velocity until no vortices are nucleated.
The bifurcation diagram observed here is similar to the one obtained in previous works \cite{HuepeCristian1997, Pham2005}, but the exact values may differ due to a different choice on the potential describing the disk.
\begin{figure}[tpb]
  \centering
  \includegraphics[width=1\linewidth,trim={0 0cm 0 0cm},clip]{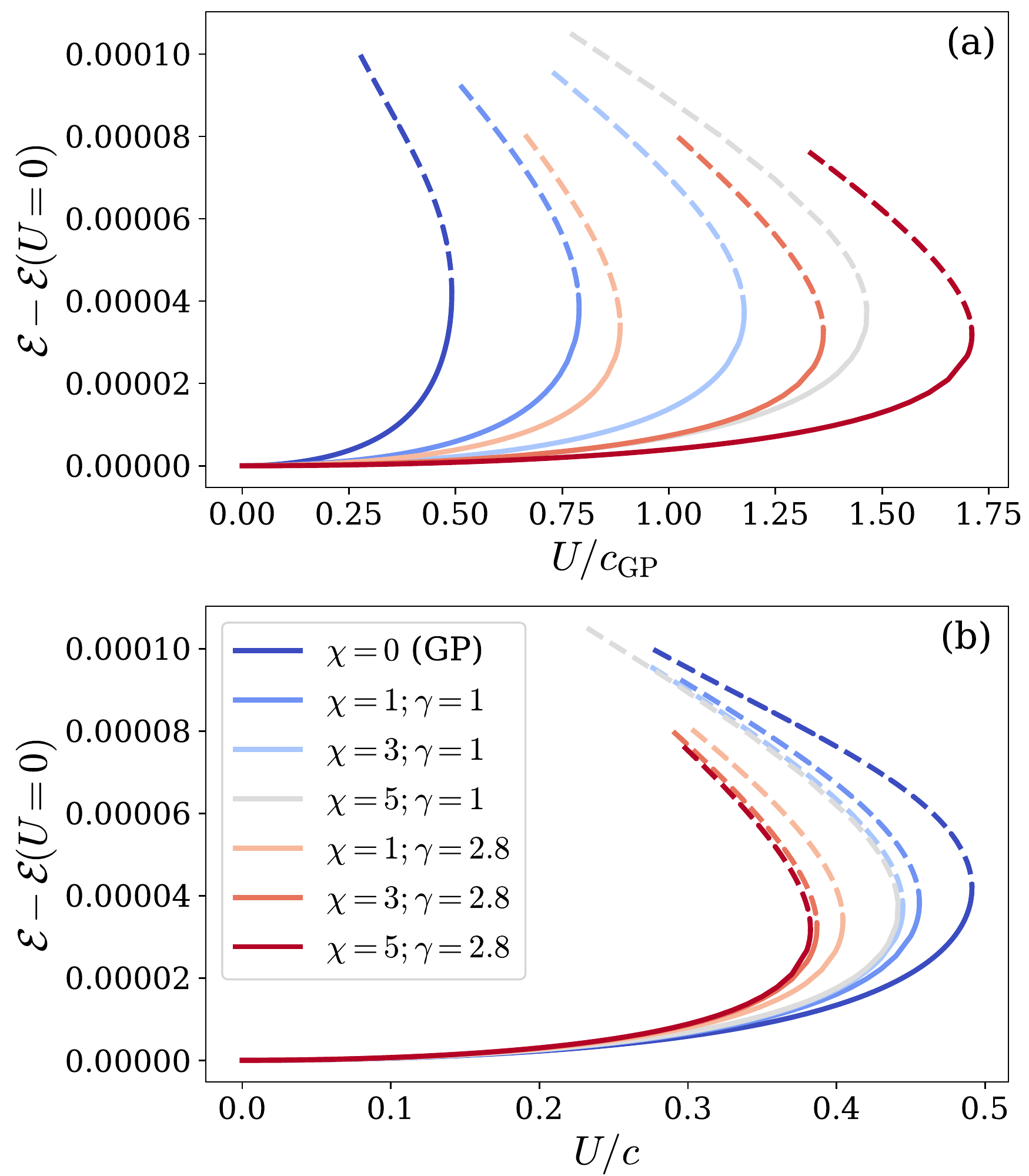}
  \caption{Bifurcation diagrams of the energy of stationary solutions of the local gGP model \eqref{eq:Enucleation} for a disk of diameter $D=5\xi$ moving at different velocities $U$. 
  The velocity is normalized by (a) the GP speed of sound $\cgp=\sqrt{gn_0/m}$ and (b) the superfluid speed of sound $c$. Different diagrams correspond to different values of the amplitude $\chi$ and order $\gamma$ of the non-linearity. The stable (solid lines) and two-vortices unstable (dashed lines) branches are shown.}
  \label{fig:d5bif}
\end{figure}%

To understand how the high-order non-linear term affects the dynamics of the system, we study the critical velocity $U_c$ of the superfluid for different values of $\chi$ and $\gamma$. Here, we use in all cases a local interaction potential $\Vxy = \delta(\bm{x}-\bm{y})$, a disk of diameter $D=5\xi$, and values of $\chi$ that go between $1$ and $5$ with $\gamma=1$ or $\gamma=2.8$. We also compare with the standard GP model ($\chi=0$). 
Note that the speed of sound and the healing length of the system depend on the values of $\chi$ and $\gamma$ according to Eqs.~\eqref{eq:c} and \eqref{eq:healing}.
In particular, we fixed in all the simulations $c=1$ and $\xi=\Delta x$. Therefore, the speed of sound varies between $c=1 \cgp$ ($\chi=0$) and $c=4.54\cgp$ (for $\chi=5$ and $\gamma=2.8$). As a consequence, the critical velocity $U_c$ in the gGP system can take relative values that are larger than $\cgp$, as shown in Fig. \ref{fig:d5bif} (a), where solid and dashed lines correspond to stable and two-vortices unstable solutions of the system, respectively. 
The increase of the relative values of $U_c$ is due to the changes on the properties of the flow, as the speed of sound of the superfluid $c$ relative to $\cgp$ increases. 
However, when the velocity is normalized by the speed of sound of the superfluid $c$, the critical Mach number $M_c$ rescales in a non-trivial manner, as shown in Fig. \ref{fig:d5bif} (b). In particular, $M_c$ decreases with the non-linearities. 
\begin{figure}[tpb]
  \centering
  \includegraphics[width=1\linewidth,trim={0 0cm 0 0cm},clip]{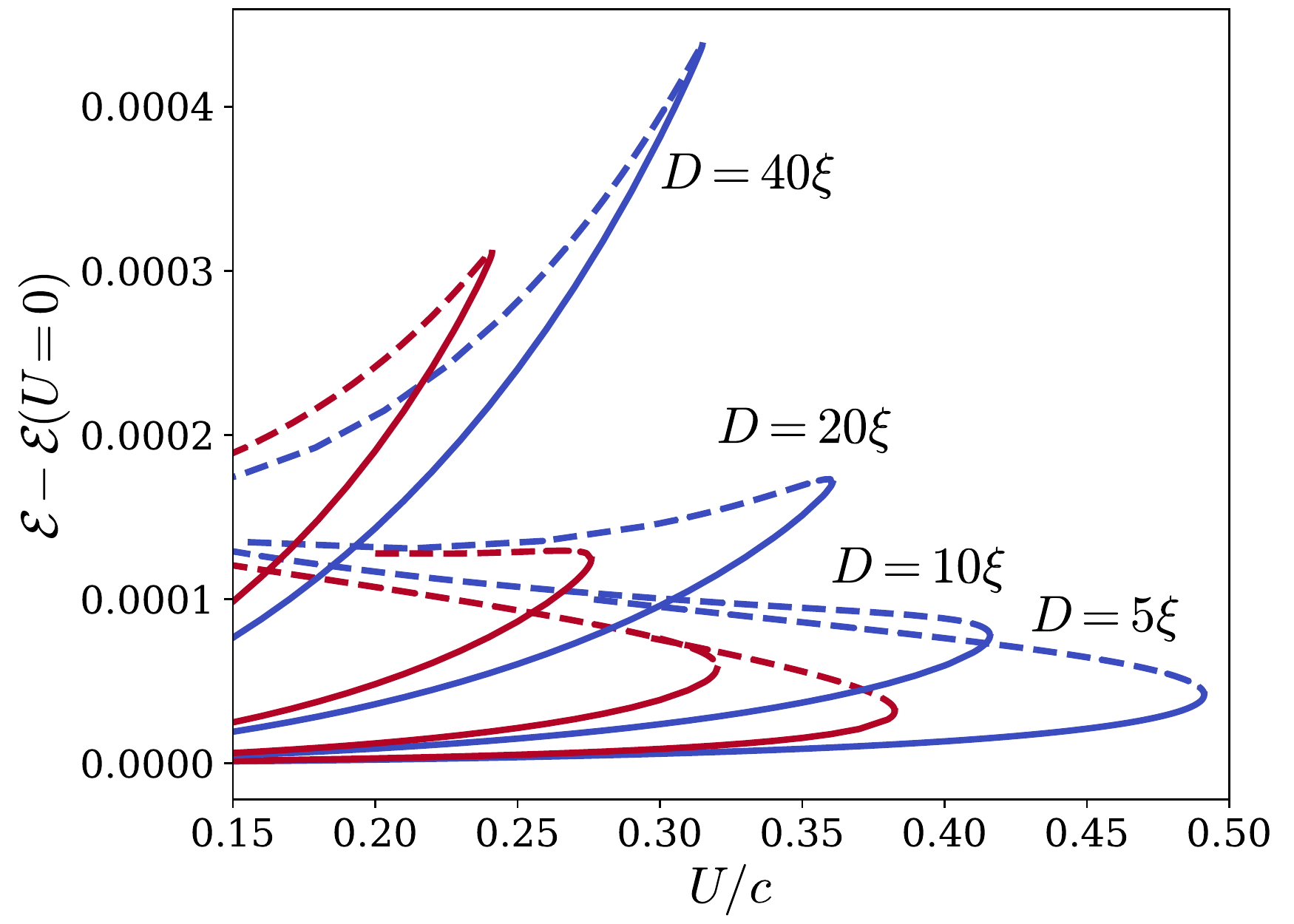}
  \caption{Bifurcation diagram of the energy of stationary solutions of a
  superfluid with a disk moving at a constant velocity $U$ for
  different diameters for the disk. Simulations of the GP model (blue lines)
  and the local gGP model with $\chi=5$ and $\gamma=2.8$ (red lines) are shown. The
  solid and dashed lines correspond to the stable and unstable branches,
  respectively.}
  \label{fig:bifur}
\end{figure}%

As already shown in Figs. \ref{fig:branches_gp} and \ref{fig:d5bif}, the critical Mach varies according to the size of the obstacle \cite{HuepeCristian1997, Sasaki2010}. 
Figure \ref{fig:bifur} shows the bifurcation diagram of a flow around a disk of diameters varying between $D=5\xi$ and $D=40\xi$. 
The blue curves correspond to the bifurcation diagram of the standard GP model ($\chi=0$) and the red curve correspond to the local gGP with $\chi=5$ and $\gamma=2.8$. Solid lines correspond to the stable branch and dashed lines to unstable solutions with two vortices. 
As the particle size $D$ increases, the critical Mach $M_c$ decreases.

\subsection{Rotons}
\label{subsec:rotons}

\begin{figure}[tpb]
  \centering
  \includegraphics[width=1\linewidth,trim={0 0cm 0 0cm},clip]{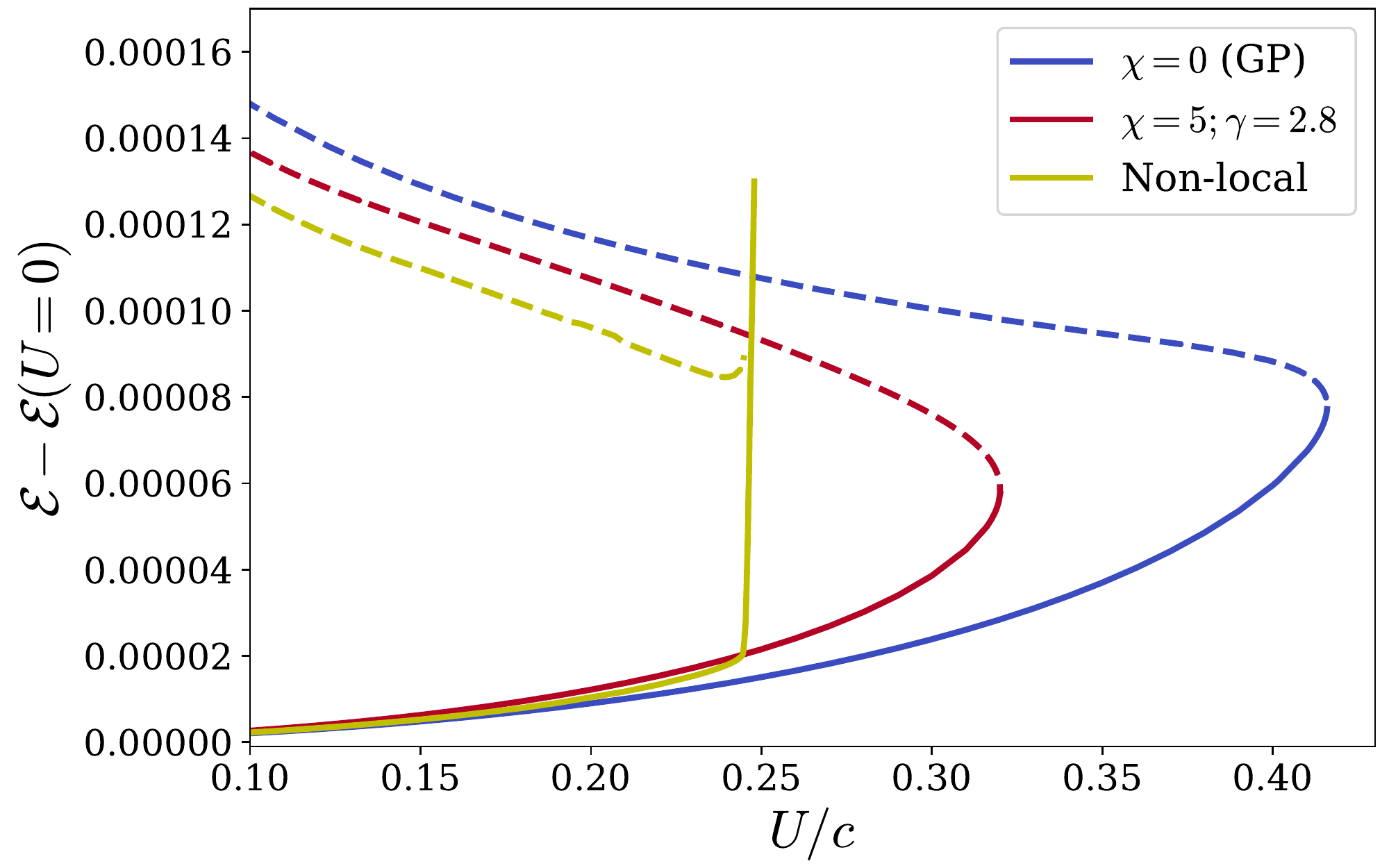}
  \caption{Bifurcation diagram of a moving disk of diameter $D=10\xi$ for a local interaction with $\chi=0$ (blue lines), $\chi=5$ and $\gamma=2.8$ (red lines) and for the isotropic non-local potential defined in \eqref{eq:potential} (yellow lines) that reproduces the roton minimum in the excitation spectrum. The solid and dashed lines correspond to the stable and unstable branches, respectively.
  }
  \label{fig:bifNLpot}
\end{figure}%

We now focus on a system with the non-local interaction potential introduced in \eqref{eq:potential}, that is able to reproduce the roton minimum in the dispersion relation \eqref{eq:bogoliubov}.
The parameters for the high-order non-linear terms are chosen as described in section \ref{sec:model}.
The bifurcation diagram of the model with rotons (yellow lines) for a disk of diameter $D=10\xi$ is shown in Fig. \ref{fig:bifNLpot}, and is compared with the local gGP (red lines) and standard GP (blue lines). The stable branch in the case with rotons presents an abrupt stop at a Mach number $M_c \approx 0.248$.
This value is close to Landau's Mach number $M_L \approx 0.245 $ obtained from applying Eq.~\eqref{eq:landau} to the dispersion relation of the gGP system \eqref{eq:bogoliubov} with the non-local potential \eqref{eq:potential} with the parameters discussed below that expression, and Landau's Mach number of \He $M_L^{\mathrm{He}} = 0.252$ assuming $c=238$ ms$^{-1}$ and $v_L = 60$ ms$^{-1}$ \cite{Donnelly1998, Godfrin2021}. 
%Note that the value is close to the Landau Mach number in superfluid helium $M_L \approx 0.25$. 
Indeed, when the disk is moving at a velocity that is larger but still close to the Landau velocity, we observe the emission of density modulations on the fluid, that can be associated with rotons. However, if the velocity of the disk is not large enough, there is no nucleation of vortices. 
This result suggests that there are two kind of excitations when a non-local interaction potential is introduced: rotons and vortices.

\begin{figure}[tpb]
  \centering
  \includegraphics[width=1\linewidth,trim={0 0cm 0 0cm},clip]{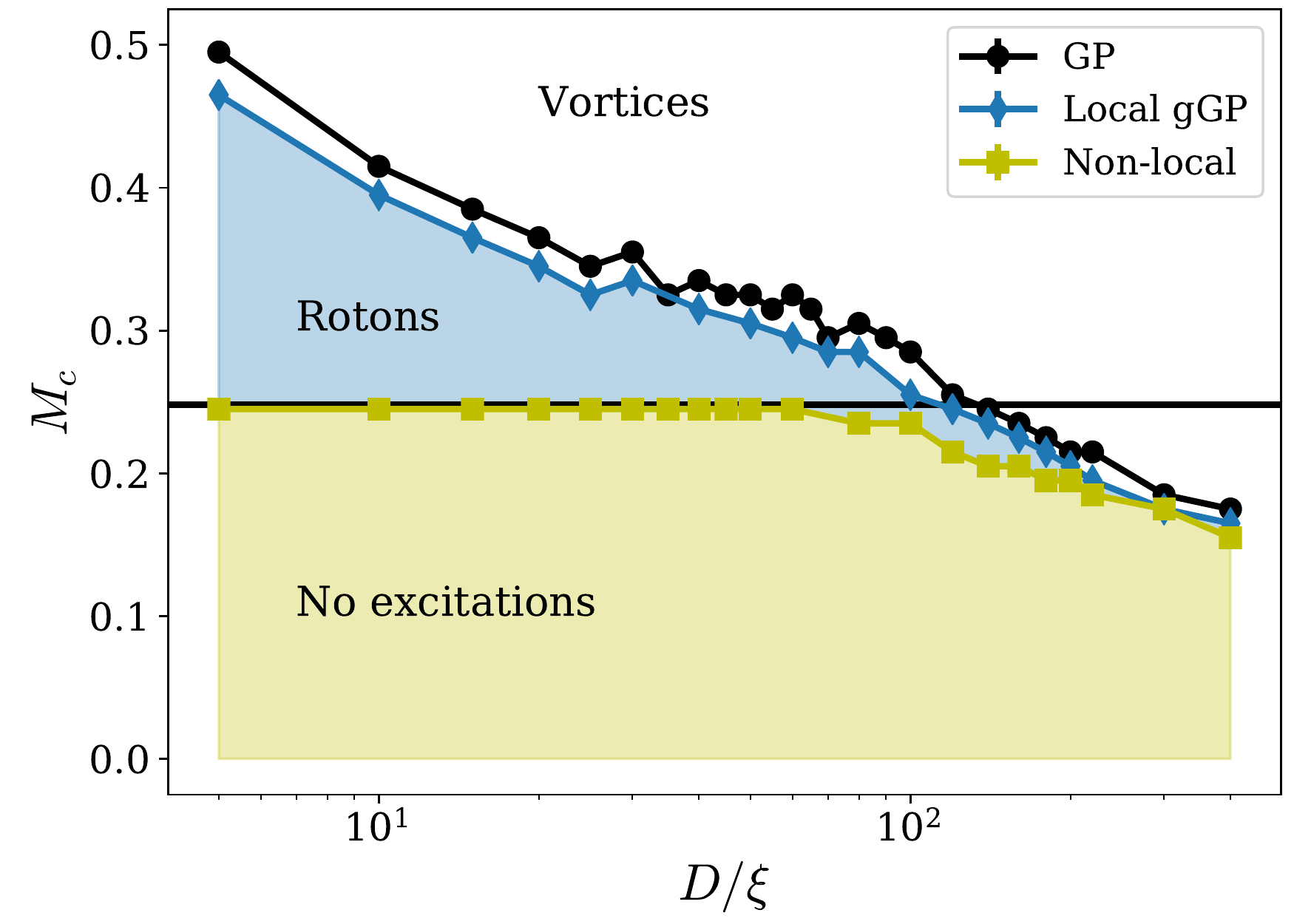}
  \caption{ Critical Mach as a function of the diameter of the disk for the standard GP model (black), local (blue) and non-local gGP (yellow) models. The horizontal solid black line indicates Landau's Mach number of the system $M_\mathrm{L}=0.248$.}
  \label{fig:critical_mach}
\end{figure}%

As discussed in section \ref{subsec:stationary}, the critical velocity for vortex nucleation depends on the size of the obstacle.
Here, we study the dependence of the critical Mach number for a wide range of disk diameters in the non-local gGP model (yellow line) and compare it with the same system with a local potential (blue lines) (Fig. \ref{fig:critical_mach}).
For comparison reasons, we also show the critical velocity dependence in the standard GP model (black line). This last one follows a similar behavior as the local gGP simulation but with larger critical values.
The system presents an interesting behavior in the case that rotons are supported.
If the disk diameter is smaller than $D\approx100\xi$, there is a range of velocities in which the disk in the non-local gGP model emits rotons but no vortices. As the diameter increases, the critical velocities for systems with and without roton minimum tend to collapse, presenting a similar behavior for large obstacles. 

According to experiments~\cite{Efimov2010}, the critical velocity in superfluid \He is of the order of $v_c \approx 10$ cm/s, value that is much smaller than Landau's velocity $v_L \approx 60$ m/s. 
However, the experiments were performed with a fork of size $D = 0.4$ mm $\approx 4\times10^6 \xi$, value four orders of magnitude larger than the largest one studied in this work of $D=400\xi$. 
The regime where only rotons are emitted would correspond to a particle size smaller than $10$ nm in superfluid \He which, to our knowledge, to this day has not been studied.
However, it is important to remark that the presence of the roton minimum seems to be irrelevant in the process of vortex nucleation for sufficiently large obstacles.

\subsection{Temporal evolution of a moving obstacle}
\label{subsec:evolution}

The solutions introduced in section \ref{subsec:stationary} provide us a better understanding of the system for the study of its temporal evolution. 
To do this, we start from a two-dimensional initial condition at rest with a disk of size $D=20\xi$ and let it evolve using the non-local gGP equation \eqref{eq:gGP} with a roton minimum in its excitation spectrum. We apply an external forcing to the particle until it achieves the desired velocity. Note that we do not include a two-way coupling in the system \cite{Giuriato2019}, i.e. the particle will not slow down after the nucleation of vortices or the emission of rotons. 
During the acceleration regime, the disk introduces small density perturbations on the flow. To mitigate spurious effects caused by these perturbations, we apply some dissipation during this regime and turn it off as soon as the target velocity is achieved.

\begin{figure}[tpb]
  \centering
  \includegraphics[width=1\linewidth,trim={0 0cm 0 0cm},clip]{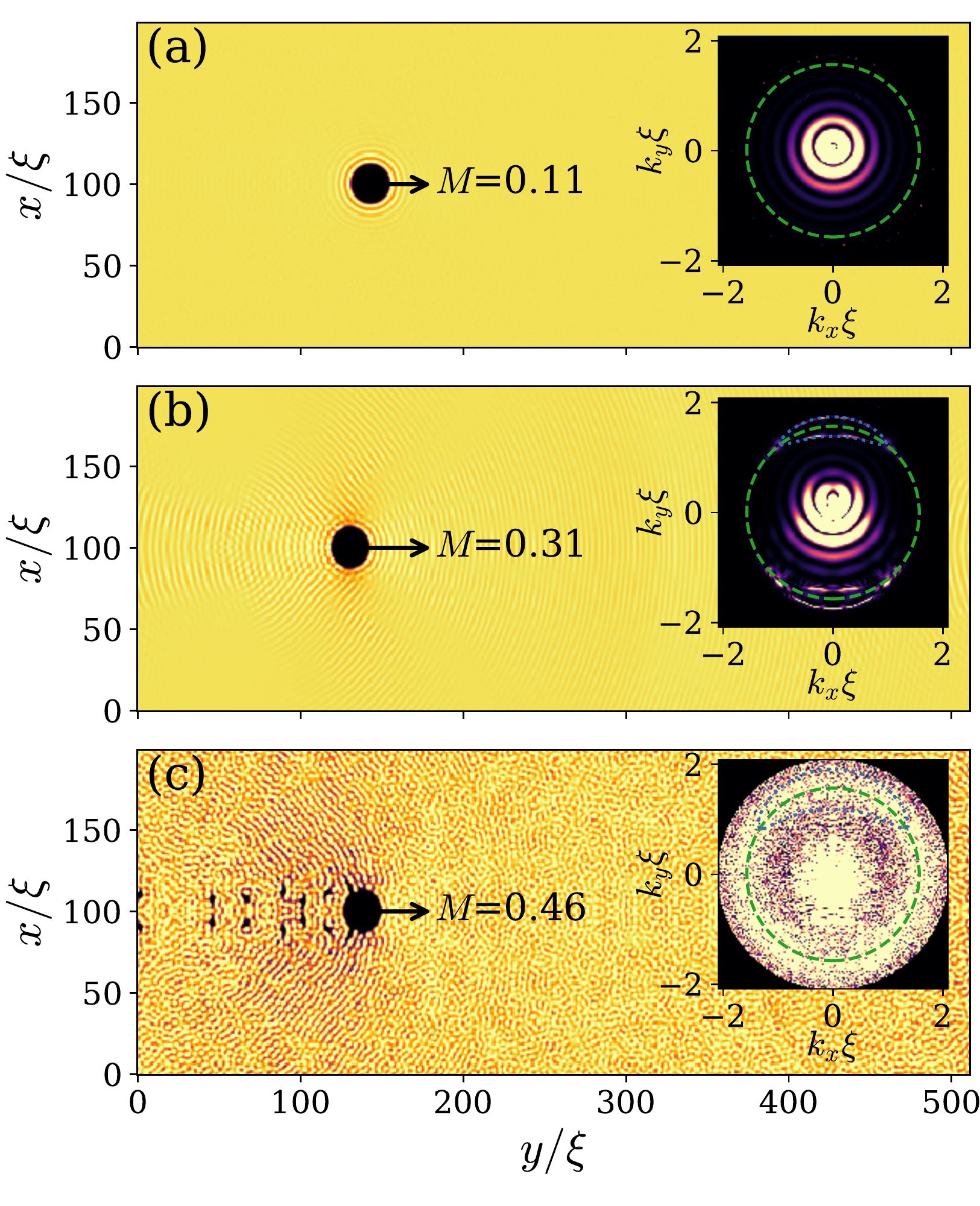}
  \caption{ Two-dimensional density fields of non-local superfluid with a disk of diameter $D=20 \xi$. 
  Dark zones correspond to regions where the superfluid is depleted. The disk is moving at a Mach number at which (a) the system is stationary, (b) rotons are emitted, and (c) vortices are nucleated, respectively. 
  The insets show the two-dimensional Fourier transform of the density field. Green dashed lines show the wave number of the roton minimum $\krot$ and the blue dotted lines solutions of Eq.~\eqref{eq:landau_vy}. 
  }
  \label{fig:evolution}
\end{figure}%

Previous works have already studied the dynamical process of vortex nucleation in the standard GP model either in two-dimensional \cite{HuepeCristian1997, Pham2005} or three-dimensional systems \cite{Winiecki2000}, observing the regular or irregular emission of vortices at the wake of the moving obstacle \cite{Sasaki2010, Stagg2014, Reeves2015}.
Here, we will focus in the non-local gGP model and the different regimes of roton or vortex emission.
Figure \ref{fig:evolution} shows snapshots of the disk moving at different Mach numbers.
In black we show the regions where the superfluid is depleted, corresponding to either the obstacle or vortices. 
For a velocity that is smaller than Landau's velocity $\ML \approx 0.245$, there are no excitations on the flow [Fig. \ref{fig:evolution} (a)]. 
Note that there are some stationary density modulations around the disk as a consequence of the non-local interaction \cite{Berloff1999, Villerot2012}, that keep their shape around the disk as it moves without causing any drag force on it.
The inset shows the two-dimensional density spectrum $|\hat{\psi}|^2(k_x,k_y)$ of the superfluid. The excited modes correspond to the density modulations around the disk. 
These patterns have already been observed around small obstacles and vortices in previous works \cite{Pomeau1993, Reneuve2018, Berloff1999}. 
More interestingly, when the particle moves at a velocity $M\gtrsim \ML$ [Fig. \ref{fig:evolution} (b)], it introduces some density fluctuations on the superfluid. 
The excited wave numbers obey the anisotropic expression in Eq.~\eqref{eq:landau_vy} computed using the dispersion relation with rotons, shown as blue dotted lines in the inset of Fig. \ref{fig:evolution} (b). 
Finally, for a velocity $M \gg M_L$ [Fig. \ref{fig:evolution} (c)], the disk emits rotons but also it starts nucleating vortices. Due to the mutual interaction between vortices and the rotons, vortices can annihilate emitting phonons. Thus, there are a wide range of modes that are excited, as shown in the inset.

In conclusion, we show here that at velocities above Landau's critical one, the moving obstacle introduces some elementary excitations with wave numbers that obey Eq.~\eqref{eq:landau_vy}. We can thus identify these excitations with rotons. For larger velocities, the disk starts nucleating vortices, emitting rotons and other excitations in a wide range of wave numbers.

\subsection{Three-dimensional system. }

\begin{figure}[tpb]
  \centering
  \includegraphics[width=1\linewidth,trim={0 0cm 0 0cm},clip]{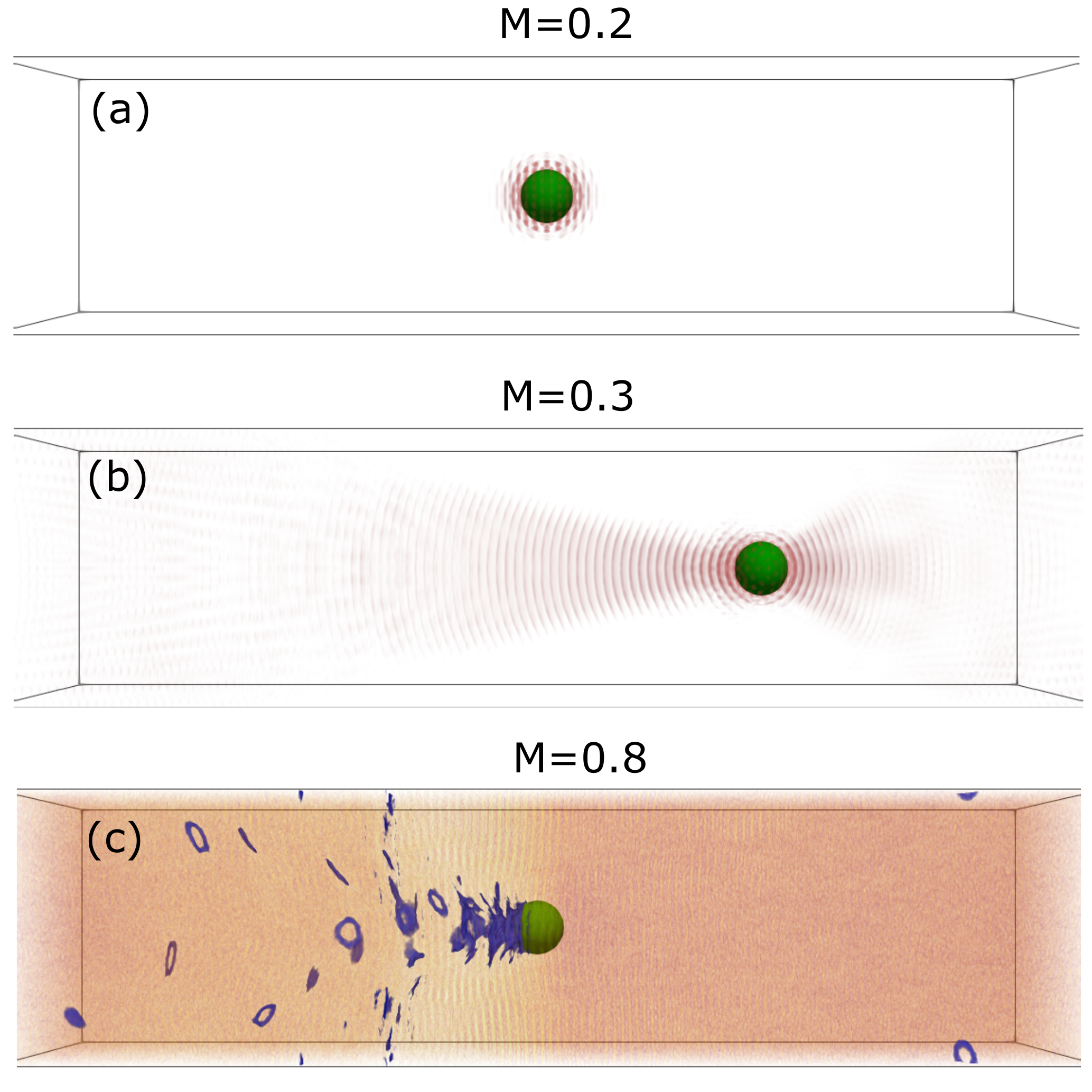}
  \caption{ Three-dimensional density field of a superfluid with a roton minimum in the excitation spectrum. 
  In green, we show a sphere of diameter $D=20 \xi$ moving to the right at Mach numbers $M=0.2$ (a), $M=0.3$ (b) and $M=0.8$ (c). 
  In red, we show density fluctuations around the equilibrium and in blue low values of the density. We can identify three different regimes, one of them stationary (a), one in which rotons in the shape of a cone are emitted (b) and one where vortices are nucleated (c). 
  %For $M=0.3$ the particles emits rotons in a cone, while for $M=0.8$ it nucleates vortices at its wake.
  }
  \label{fig:3D_evolution}
\end{figure}

All of the results discussed until now were obtained from two-dimensional simulations of the non-local gGP model. A similar behavior can be obtained in three-dimensional systems. 
In particular, we studied the motion of a sphere of diameter $D=20\xi$ in the $z$-direction in an elongated domain with $L_z=4 L_{\perp}$ and a spatial resolution of $256\times256\times1024$. 
 (a)]. The density fluctuations in red around the sphere create a pattern induced by the roton minimum but that do not emit any excitation on the flow.
In the case of the sphere moving at a velocity $M=0.3$ just above Landau's Mach number, it starts emitting rotons in the shape of a cone, shown as red density fluctuations above the equilibrium in Fig. \ref{fig:3D_evolution} (b). Note that the cone is emitted in both upstream and downstream directions, consistent with negative solutions in the wave numbers shown in the inset of Fig. \ref{fig:evolution} (b). 
For a larger velocity $M=0.8$, the particle starts nucleating vortices (blue rings in Fig. \ref{fig:3D_evolution} (c)). 
The depletion of the superfluid is stronger at the wake of the sphere, where vortices are nucleated. In this region, many vortex rings reconnect and collapse due to the interaction with strong density fluctuations introduced by the roton minimum, shown in red.

\begin{figure}[tpb]
  \centering
  \includegraphics[width=1\linewidth,trim={0 0cm 0 0cm},clip]{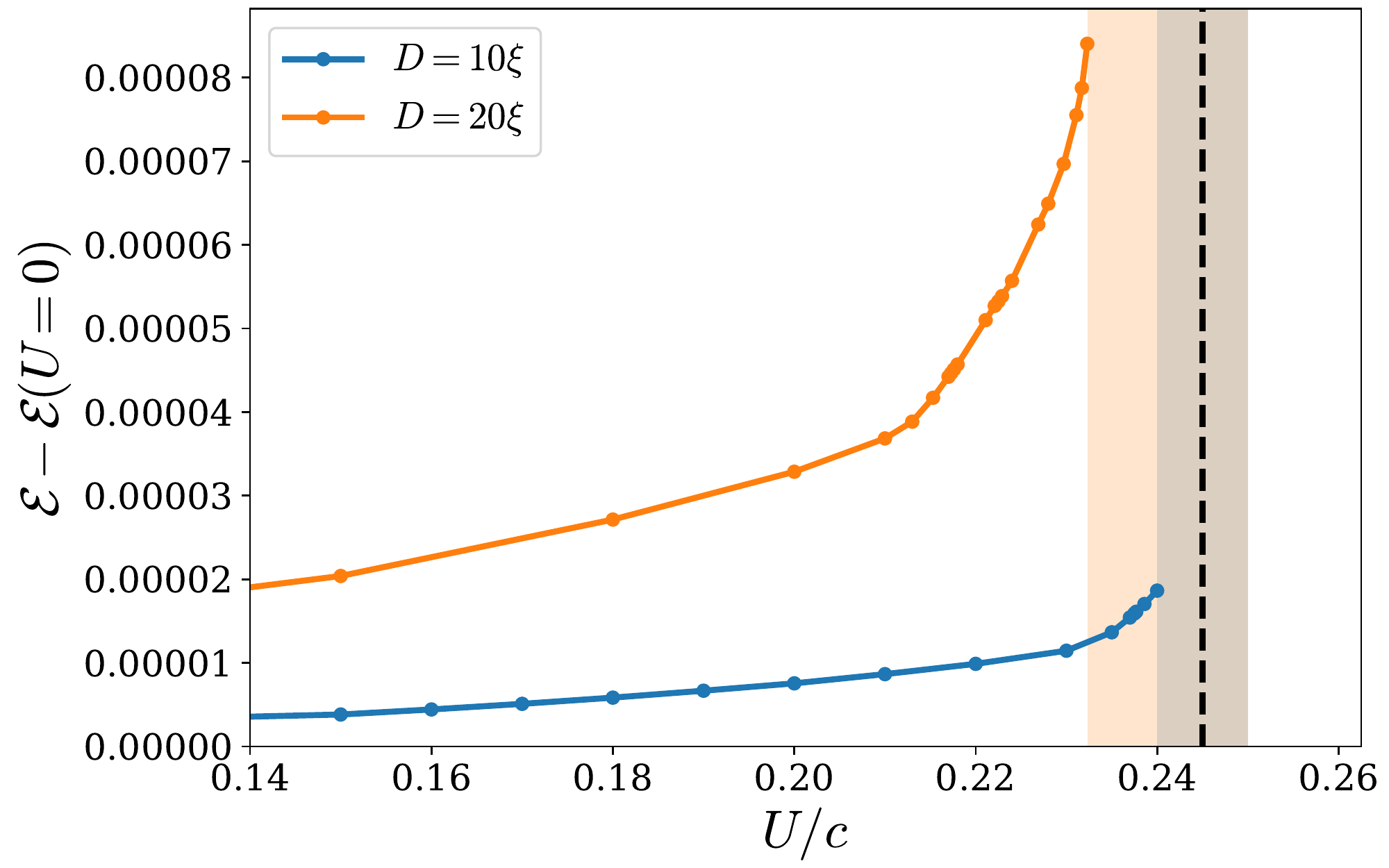}
  \caption{ Stable stationary solutions of a moving sphere of two different diameters $D$. The highlighted regions correspond to the intervals where we estimate the critical Mach number should be found, with the lower bound being the maximum value obtained using the Newton-Raphson method and the upper bound obtained using the imaginary time evolution of Eq.~\eqref{eq:gGP_disk}.
  Vertical black line indicates Landau's Mach of the system $\ML=0.245$. 
  }
  \label{fig:critical_3D}
\end{figure}

We also perform an analysis on the critical velocity in the three-dimensional case for two sphere diameters $D=10\xi$ and $D=20\xi$. These sizes correspond to the small particle limit discussed in the two-dimensional case and are chosen in this way to avoid spurious effects introduced by the boundary conditions. Larger particle sizes require larger computational boxes that are prohibitive.  
The critical velocity of the system can be determined by the Mach number where the stable and unstable branches merge. The unstable branch can only be obtained using a Newton-Raphson method that is too expensive in three-dimensions and is out of the scope of this work. Therefore, we only show the stable branch in figure \ref{fig:critical_3D} for both particle sizes.
The stable branch allows us to determine a lower bound of the critical Mach value, corresponding to the maximum value of $M$ at which the Newton-Raphson method converges. 
We have checked that the imaginary time evolution of the gGP model, obtained by replacing $t \rightarrow -it$ in Eq.~\eqref{eq:gGP_disk}, does not converge for $M=0.25$, which is slightly above the theoretical value for roton emission $\ML=0.245$. Therefore, we can estimate that the critical Mach belongs to the interval $M_c \in [0.24,0.25]$ for $D=10\xi$ and $M_c \in [0.231, 0.25]$ for $D=20\xi$ (highlighted regions in Fig. \ref{fig:critical_3D}). 
These results are consistent with the ones obtained in two dimensions for small particles $D < 100\xi$. 
A more precise determination of the critical value in three dimensions and the study of the unstable branch is left for a future study.

\section{Conclusions}
\label{sec:conclusions}

In this work, we study the process of vortex nucleation at the wake of a moving obstacle in a generalized Gross-Pitaevskii (gGP) model \cite{Muller2020,Gallemi2020a} in periodic two-dimensional systems. 
We determine the critical velocity of the superfluid, velocity above which superfluidity breaks down, for moving disks of diameters between $D=2.5\xi$ and $D=400\xi$ by analyzing the bifurcation diagram of stationary solutions of the system \cite{HuepeCristian1997, Pham2005}.
%All the results are compared with the standard GP model, 
In particular, we study the role of the beyond mean field corrections and the introduction of a non-local interaction potential that can reproduce the roton minimum in the excitation spectrum, observed in superfluid \He and in dipolar BECs \cite{Godfrin2021,Lahaye2009}, and compare them with the standard GP model.

Varying the amplitude and order of the high-order non-linear terms in the local gGP model, we show that the role of beyond mean field corrections is to reduce compressible effects in the system, increasing the value of the speed of sound $c$ and decreasing the core size of the vortices. As the absolute value of the speed of sound increases, the critical velocity also does. However, it does not do it in a trivial way as the critical Mach number decreases with the non-linearities. 

In the case of a non-local interaction potential, we show that the superfluid presents two characteristic velocities, one of them associated with the emission of rotons and the other related with the vortex nucleation. 
In the case of impenetrable disks of diameter $D \lesssim 100 \xi$, the critical velocity is a consequence of the roton minimum in the excitation spectrum. Above $\ML\approx 0.25$, the disk starts emitting rotons that, in the case of the particle moving in the $y$-direction, satisfy the dispersion relation \eqref{eq:landau_vy}. For small obstacles, there is a range of velocities where only rotons are emitted and no vortices are nucleated. In this case, rotons are the reason for the break down of superfluidity.
For larger obstacles $D \gtrsim 100\xi$, the critical velocity for nucleation of vortices becomes smaller than the one for emission of rotons, and its value for the different models tend to collapse, suggesting that for large obstacles the rotons are not relevant in the mechanism of vortex nucleation.
In the case of a three-dimensional system, the dynamics of a moving sphere immersed in a superfluid is consistent with the behavior observed in the two-dimensional case, although the study is limited to small particles because of computational constraints.
We identify the presence of three regimes for different particle velocities, one in which the moving particle does not perturb the flow, one in which it emits rotons and a third one in which it nucleates vortices perturbing the flow in the whole range of scales. The critical velocity in the limit of small particles is consistent with Landau's critical velocity of the system.
%\DEL{In the case of a three-dimensional system, the dynamics of a moving sphere immersed in a superfluid follows the same behavior as in the two-dimensional case. In particular, for a sphere of diameter $D = 20\xi$, the critical velocity for vortex nucleation is larger than the one of roton emission. }

\begin{acknowledgments}
  This work was supported by the Agence Nationale de la Recherche through the project GIANTE ANR-18-CE30-0020-01. This work was granted access to the HPC resources of CINES, IDRIS and TGCC under the allocation 2019-A0072A11003 made by GENCI.
  Computations were also carried out at the Mésocentre SIGAMM hosted at the Observatoire de la Côte d'Azur.
\end{acknowledgments}

\appendix

\section{ Newton–Raphson method}
\label{sec:newton}

In order to find the critical velocity at which a superfluid breaks down, i.e. the velocity above which vortices are nucleated, we can study stationary solutions of the system of either maximum or minimum of the energy $\mathcal{E}$ \eqref{eq:Enucleation}. 
One way of doing this is to study the imaginary time gGP, obtained by replacing $t \rightarrow -it$ in Eq.~\eqref{eq:gGP_steady}. 
%To do this, the generalized Advective Real Ginzburg-Landau (gARGL) equation \cite{Nore1997} associated to \eqref{eq:gGP_vel} (obtained by replacing $i\rightarrow  1$) may be studied. 
The evolution of this equation allows one to obtain a ground state of the system, that corresponds to a stable stationary solution of the system.
%advected gGP model \eqref{eq:gGP_vel} with minimal energy. 
However, for the vortex nucleation problem we expect to find a bifurcation diagram with stable and unstable solutions of the gGP equation \cite{HuepeCristian1997,Pham2005}, so this method would only allow us to obtain the stable branch of the system.

An alternative way of computing the stationary solutions of \eqref{eq:gGP_steady} is by using the  Newton–Raphson method \cite{Tuckerman2004}. To find both, stable or unstable steady states, we study an equation of the form
\begin{equation}
\frac{\partial \Psi}{\partial t} = 0 =L\Psi + N(\Psi) + A(\Psi),
\label{eq:LW}
\end{equation}
where $L$ correspond to a linear operator, $N(\Psi)$ is an arbitrary function involving multiplicative and non-linear terms, and $A(\Psi)$ corresponds to the advective term. The Newton–Raphson method consists in finding iteratively a solution of the above problem. We start from an initial guess $\Psi$, which is then perturbed as $\Psi-\delta\psi$, with $\delta\psi$ small. By linearizing equation \eqref{eq:LW} for small $\delta\psi$, we obtain the following linear equation
\begin{equation}
(L+DW(\Psi))\delta\psi = L\Psi + W(\Psi),
\label{eq:linLW}
\end{equation}
with $DW(\Psi)$ the Jacobian of $W(\Psi) = N(\Psi) + A(\Psi)$ at $\Psi$, acting on $\delta\psi$. To solve numerically this equation, we use an iterative bi-conjugate gradient stabilized method (BiCGSTAB) with a preconditioner $\mathcal{P} = (I - \Delta t L)^{-1}$, where $\Delta t$ is an arbitrary parameter used to improve convergence \cite{Tuckerman2004}.
Newton–Raphson method can only be used when a good estimation of the steady state is provided as initial guess.

%--------------------- ----------------------

\bibliography{Nucleation}

\end{document}